# Water/Cosolvent Attraction Induced Phase Separation: a Molecular Picture of Cononsolvency

Taisen Zuo[1,2], Changli Ma[1,2], Guisheng Jiao[1,2], Zehua Han[1,2,3], Shiyan Xiao[4], Haojun Liang[4], Liang Hong[5], Daniel Bowron[6,*], Alan Soper[6,†], Charles C. Han[7], He Cheng[1,2,‡]

[1] China Spallation Neutron Source (CSNS), Institute of High Energy Physics (IHEP), Chinese Academy of Science(CAS), Dongguan 523803, China

[2] Dongguan Institute of Neutron Science (DINS), Dongguan 523808, China

[3] University of Chinese Academy of Sciences, Beijing 100049, China

[4] CAS Key Laboratory of Soft Matter Chemistry, Collaborative Innovation Center of Chemistry for Energy Materials, Department of Polymer Science and Engineering, University of Science and Technology of China, Hefei, Anhui 230026, China.

[5] School of Physics and Astronomy &Institute of Natural Sciences, Shanghai Jiao Tong University, Shanghai 200240, China

[6] ISIS Facility, Rutherford Appleton Laboratory, Harwell Science and Innovation Campus, Didcot OX11 0QX, United Kingdom

[7] Institute for Advanced Study, Shenzhen University, Shenzhen, 508060, China

Cononsolvency is a phenomenon for which the solubility of a macromolecule decreases or even vanishes in the mixture of two good solvents. Although it has been widely applied in physicochemical, green chemical and pharmaceutical industry, its origin is still under active debate. Here, by using combined neutron total scattering, deuterium-labelling and all-atom molecular dynamic simulations, we demonstrated that it is the strong water/cosolvent attraction that leads to the cononsolvency. The combined approach presented here has opened a new route for investigating the most probable all-atom structure in macromolecular solutions and the thermodynamic origin of solubilities.

Tel: +86-769-8915-6445; Fax: +86-769-8915-6441.

The solubility of macromolecules is of fundamental importance in many scientific disciplines and industrial applications. Unfortunately, the existing theoretical framework to describe the solvation of macromolecules are constructed primarily based on empirical observations without knowing the real solubility parameters, which cannot be measured directly[1,2]. One problematic consequence of this inability to directly measure the real macromolecular solubility parameter, is that several important dissolution phenomena cannot then explained in the existing framework of understanding; cononsolvency is a typical example[3].

Since the first evidence of cononsolvency was found in the 1980s, theoretical debates on its origin have continued. Currently, there are four main hypotheses, i.e., the perturbation of the water-cosolvent interaction with the presence of polymer network[4], Competitive adsorption[5,6], the formation of a stoichiometric complexation between water and cosolvent[7], and strong water-cosolvent interactions[8]. Although still widely considered, we note that the first hypothesis was tarnished when Schild et al.[9] found that similar cononsolvency phenomena happened in both macromolecular dilute solution and gel. The second one focuses on the cooperative competition between solute-water and solute-cosolute hydrogen bonding or the strong solute-cosolvent interaction[10]. This assumption assumes that the water-cosolvent interaction is too weak to be taken into account. The formation of a stoichiometric water-cosolvent complexation is the third hypothesis[7, 11], where it is proposed that the complexation can be considered to result in a new "compound" which is a poor solvent for the macromolecule. Recently, we ourselves pointed out that thermodynamics still plays an important role, and that a strong water/cosolvent interaction could be the origin for the cononsolvency phenomenon. Small angle neutron scattering (SANS) was used to investigate the phase behaviour of a macromolecule in a mixed solvent, and three interaction parameters, i.e., macromolecule-water, macromolecule-cosolvent and water-cosolvent were analysed according to the ternary Random Phase Approximation model[3, 8, 12, 13]. Dudowicz et al. refined the classic Flory-Huggins theory to consider the mutual association of the solvent molecules, and they suggested that a large negative solvent-cosolvent interaction parameter should be a necessary condition for the occurrence of cononsolvency[14, 15].

In spite of the large body of existing experimental work, puzzles remain because none of the former experiments has been able to directly observe the solvation of the macromolecules in mixed solvents at the atomistic level directly. All of the previous conclusions were made simply based on experimental observation of either the phase diagram, or the rough conformation variations of macromolecules[3, 16].

Recent developments in Neutron Total Scattering techniques facilitate the continuous structural measurements covering the length scale from 0.01 angstrom to 10 nanometres[17]. The new instrumentation NIMROD[18] when combined with deuterium-labelling techniques and molecular dynamic (MD) simulation allows us to connect the most-probable all-atom positions of the mixed solvent with the macromolecule conformation. Targeting the puzzle of cononsolvency, we have carried out the measurement and analysed the scattering profiles of poly(*N*-diethylacrylamide) (PDEA) water-ethanol solutions with different deuterium ratios and at different ethanol concentrations in the one phase region, away from the phase boundary, and deduced the most-probable spatial distribution of each atom. These results demonstrate that cononsolveny can be explained in a framework of equilibrium thermodynamics with fluctuations. The strong water/cosolvent (water-ethanol) attraction is the necessary condition, while both preferential adsorption and water-ethanol complexation are results of the intermolecular interactions of the thermodynamic laws.

Following the phase diagram of PDEA in water-ethanol solution as shown in Fig. 1(a) [12], a series of neutron total scattering experiments were conducted. PDEA has a lower critical solution temperature (LCST) phase diagram in the ethanol-poor region, and is totally soluble (the phase boundary cannot be detected) in the ethanol-rich region. Four ethanol concentrations in the one phase region were chosen for the neutron total scattering experiments at 20 ℃. To extract further insight into the structural correlations in the samples, a series of isotopically labelled solutions was studies for each concentration point: PDEA/$D_2O$/$C_2D_5OD$ (Fully Deuterated), PDEA/$H_2O$/$C_2H_5OH$ (Fully protonated), and PDEA/HDO/$C_2M_5OM$ (Half Deuterated, M means half D and half H). In total this resulted in 12 samples measured as shown in Fig. S1. All-atom MD simulations were conducted at all of the four ethanol concentrations, as shown in Fig. 1(a). The simulation boxes are about 17x17x17 nm$^3$ containing ~520000 atoms(Fig. 1(c)) to keep the number density of each atom type the same as the neutron scattering experiment. The MD-derived neutron scattering profiles are in a quantitative agreement with the experimental data (see Fig. 1(b)), confirming that both the conformation of PDEA and the spatial distributions of small molecule observed in experiment have been reasonably represented by our MD simulation models.

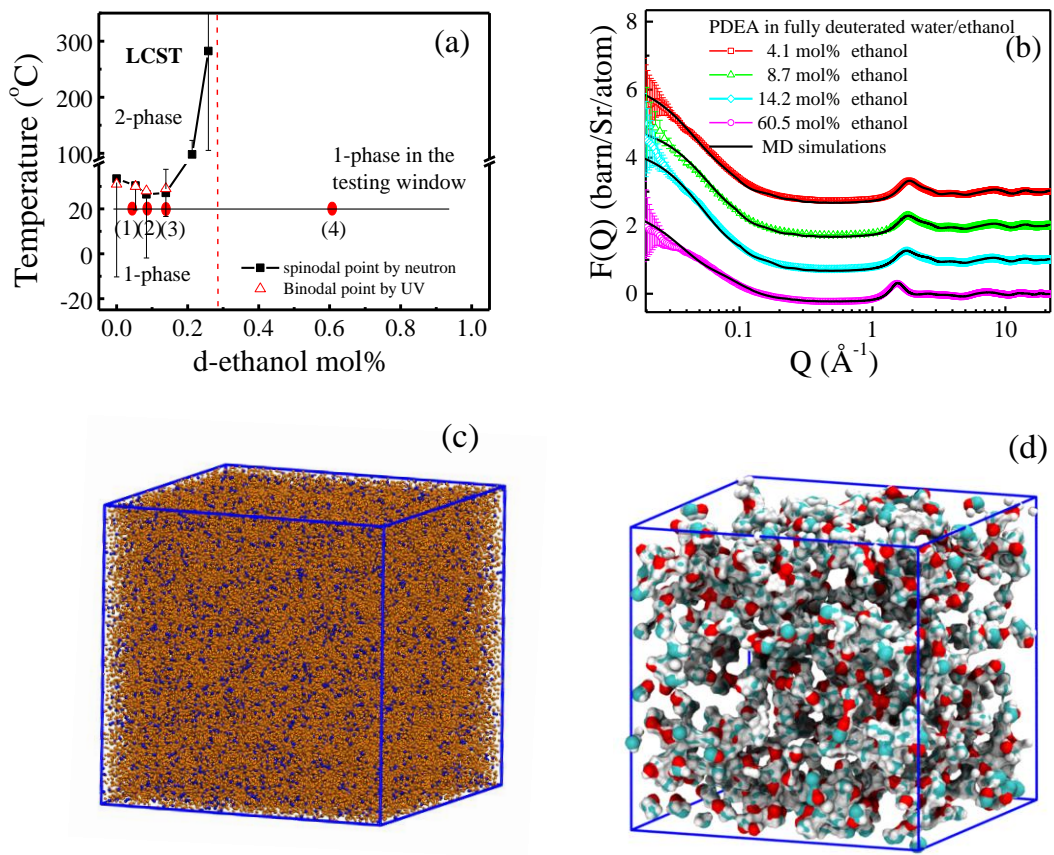

FIG. 1. Neutron total scattering and MD simulations. (a) The phase diagram of PDEA in water-ethanol solution from our previous work of small angle neutron scattering experiments[12]. The neutron total scattering experiments in this work were conducted at 20 ℃ at different ethanol concentrations, e.g., (1) 4.1 mol%, (2) 8.7 mol%, (3) 14.2 mol% and (4) 60.5 mol%, respectively. (b) The neutron total scattering profiles of PDEA in fully deuterated water-ethanol solutions at different ethanol concentrations. The black solid lines are the neutron scattering profiles derived from MD simulations. (The scattering and calculated curves of 4.1 mol%, 8.7 mol% and 14.2 mol% ethanol were shifted upward 3, 2 and 1 barn/Sr/atom, respectively). (c) The simulation box of PDEA in water-ethanol solution at 8.7 mol% ethanol. The blue dots are ethanol molecules, and the orange dots represent water molecules. (d) The snap shot of the solvent accessible surface of the ethanol molecules in the simulated box of 8.7 mol% ethanol (The size of the box was 6×6×6 nm$^3$ which was cut from the big simulation

box shown in Fig. 1(c). Only the ethanol molecules were shown in the snapshot where the red balls are oxygen atoms, the blue balls are carbon atoms and the white balls are hydrogen atoms).

The origin of cononsolvency can be explored according to the most probable all-atom positions as shown in Figs. 1(c) and (d). Let's examine the four leading explanations one by one: the first explanation proposed by Amiya et al.[4] is that the re-entrant phase transition occurs because the attractive interaction between alcohol and water is enhanced by the presence of macromolecules. The neutron total scattering profiles of the water-ethanol mixture with or without PDEA show no differences as shown in Fig. S5. The resultant partial pair distribution functions further support this point (Fig. 2 and Fig. S6). The liquid water is considered as a tetrahedrally coordinated random network[19]. On average, 3.5 water molecules are located around a central one[20]. When the ethanol concentration is lower than 14.2 mol%, the positions of the first two peaks in water-water pair distribution function (Fig. 2(a)) keep constant, indicating that the structure of mixed solvent tends to retain its tetrahedral water structure. This is consistent with neutron scattering results in aqueous alcohol solutions by Soper et al.[21, 22] and MD simulations by Fidler et al.[23]. When ethanol concentration increases to 60.5 mol%, the tetrahedral structure is distorted, i.e. the second peak in g(r) $_{OW-OW}$ shifts from 4.5 Å to 4.8 Å. At this high concentration, the structure of the mixed solvent tends to be the zigzag structure of pure ethanol[24], as shown in Fig. 2(b). The average nearest ethanol oxygen-oxygen distance is 2.8 Å, and the second nearest ethanol oxygen-oxygen distance is 4.8 Å (inset in Fig. 2(b). Similar structure variation of alcohol water mixtures had been observed by Yamaguchi et al. [25]. Taken together, these findings suggest that the presence of macromolecules in moderate concentration does not perturb the structure of the mixed solvent.

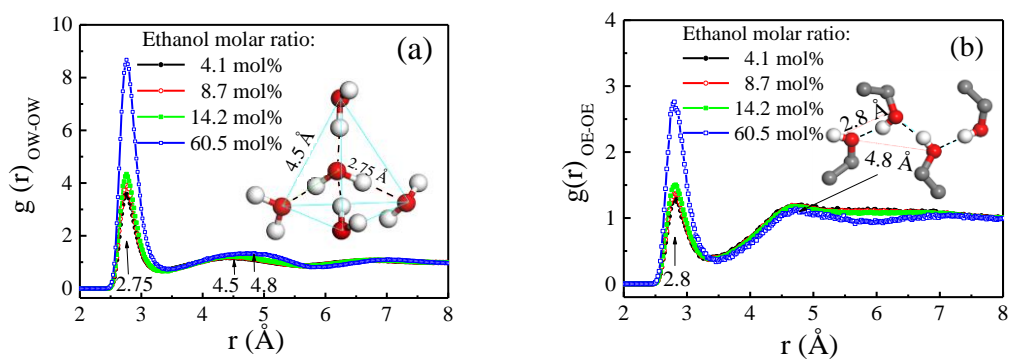

FIG. 2. Local structure of water (W) and ethanol (E). (a) g(r) $_{OW-OW}$ (OW means oxygen atoms of water, HW is the hydrogen atom of water), the inset is the tetrahedral structure of liquid water; (b) g(r) $_{OE-OE}$, the inset is the zigzag structure of ethanol (OE means oxygen atom of ethanol, HE is the hydroxyl hydrogen atom of ethanol).

The second explanation is preferential adsorption. There are two types of preferential adsorption proposed by two different groups. The first posed by Tanaka et al.[10, 26] was based on the competitive hydrogen bonding of water and alcohol to the macromolecule, while the second, suggested by Mukherji et al.[27] was due to the strong interaction between the alcohol and the macromolecule. Mukherji et al. applied an adaptive resolution scheme (AdResS) method with a Metropolis particle exchange criterion to the re-entrant behaviour of Poly(*N*-isopropylacrylamide) (PNIPAM) in water-methanol solution, and concluded that the preferential solute-cosolvent interaction is the key[27]. However, from the combined neutron total scattering and MD simulations, we found that preferential adsorption is just a result of cononsolvency when the ethanol concentration is lower than 60.5 mol%.

It is well known that a macromolecule needs a hydration layer on its surface to be dissolved in water (Fig. 3(a)). So, averaged pair distribution functions of both the ethanol oxygen (OE) and the water oxygen (OW) from the carbonyl groups of PDEA (proton acceptor) were plotted. The first peak at about 2.74 Å and 2.77 Å in Figs. 3(b) and 3(c) clearly demonstrates the hydrogen bonding of ethanol and water with the carbonyl group of PDEA. The length of the average hydrogen bonding of ethanol-PDEA is longer than that of water-PDEA, which proves that the former interaction is relatively weaker. The first peak of $g(r)_{C=O \cdots OE}$ decreases while that of $g(r)_{C=O \cdots OW}$ increases with ethanol concentration. Note that the first and fourth peaks in $g(r)_{C=O \cdots OE}$ are higher than 1 at 4.1 mol% ethanol concentration, then all of $g(r)_{C=O \cdots OE}$ start to be lower than 1 with the increase of ethanol concentration (Fig. 3(b)). It indicates that the dynamical local density of ethanol molecules inside the hydration layer is higher than its average value in the system at 4.5 mol% and 8.7 mol% ethanol concentrations, then it becomes lower than the average ethanol concentration in the system at higher ethanol concentrations. On the other hand, except the height of the first peak in $g(r)_{C=O \cdots OW}$ at 60.5 mol% ethanol concentration, all of $g(r)_{C=O \cdots OW}$ are lower than 1. It shows that, the dynamical water concentration inside the hydration layer is higher than that in the system at 60.5 mol% ethanol concentration (Fig. 3(c)). Therefore, both competitive hydrogen bonding of ethanol/water with PDEA and preferential adsorption of ethanol[28] on the surface of PDEA have been observed in our system. However, the preferential adsorption of ethanol happens only when ethanol concentration is lower than 60.5 mol%, and it is not the origin of cononsolvency[29].

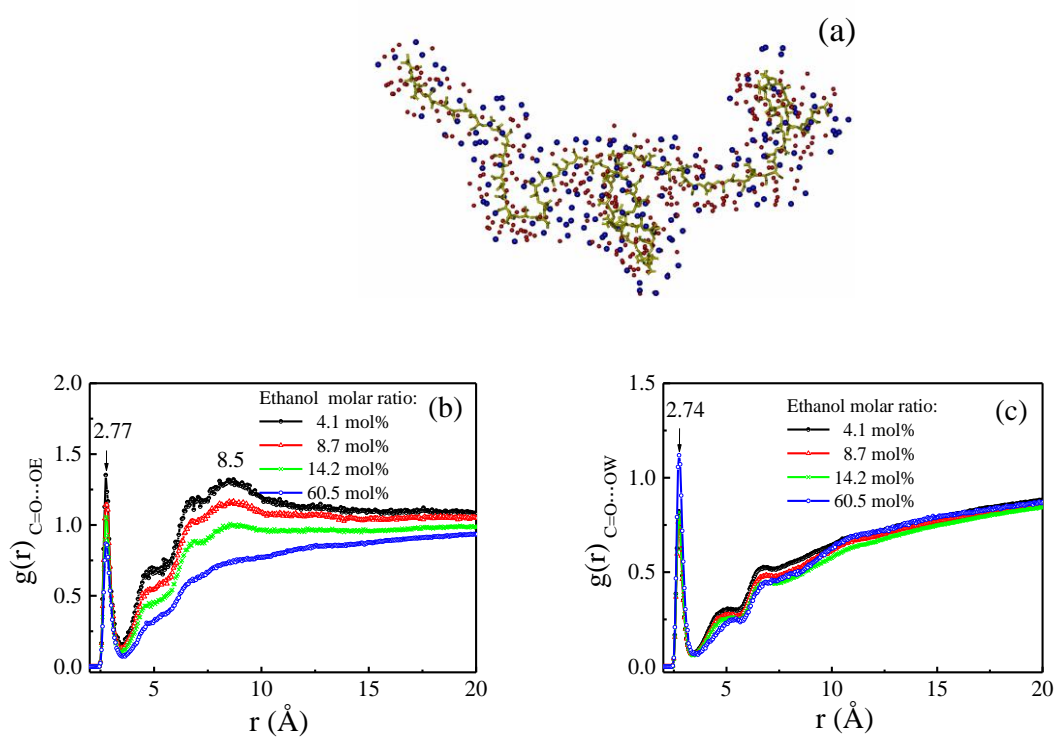

FIG. 3. Competitive hydrogen bonding of ethanol/water on the carbonyl group and preference absorption of ethanol on the surface of PDEA. (a) The snap shot of PDEA in the simulation box and the first solvation layer of water and ethanol around it (the red dots are water molecules and the blue dots are ethanol molecules). (b) The partial pair distribution function $g(r)_{C=O \cdots OE}$ of the oxygen of ethanol around the carbonyl oxygen atom of PDEA at different ethanol molar concentrations, e. g., 4.1 mol%, 8.7 mol%, 14.2 mol% and 60.5 mol%, respectively. (c) Partial pair distribution function $g(r)_{C=O \cdots OW}$ of the oxygen of water around the carbonyl oxygen atom of PDEA at different ethanol molar concentrations, e. g., 4.1 mol%, 8.7 mol%, 14.2 mol% and 60.5 mol%, respectively.

The formation of a stoichiometric complexation is the third explanation. Zhang et al. used static and dynamic laser light scattering to investigate the coil-to-globule-to-coil re-entrance in PNIPAM water-methanol solution. Because the compact globule state of PNIPAM appears when methanol concentration reaches 17 mol%, they suggested that a 5:1 stoichiometric complexation between water and methanol should exist[11]. Here, we can visualize "complexations" in water ethanol solution directly in Figs.1(c) and 1(d).

Heterogeneity exists in PDEA/water/ethanol mixtures, and they are not stoichiometric. Dixit et al. studied the concentrated alcohol–water mixture (7:3 molar ratio) by neutron total scattering and also found an incomplete mixing or complexation at molecular level[30].

Cononsolvency can happen spontaneously because of the negative Gibbs free energy change. The ternary system either gives off energy ($\Delta H < 0$), or becomes more disordered ($\Delta S > 0$), or both. Let's first examine the water-ethanol interaction. Experimentally, the enthalpies of water-ethanol mixtures had been measured at 20 ℃ in the 1970s as shown in Fig. 4(a)[31]. The MD simulations enable us to calculate the excess enthalpies at different ethanol concentrations with or without PDEA. Simulation agrees well with the experimental data at low ethanol concentration up to 14.2 mol% ethanol, and deviates at high ethanol concentration. This might originate from the inaccuracy of the OPLS force field of ethanol[32]. The negative excess enthalpy of water-ethanol mixtures indicates the strong water-ethanol attraction. It explains the facts that ethanol is totally miscible with water and the azeotropic temperature is 78.2 ℃ [33]. There is almost no change in the excess enthalpy, as shown in Fig. 4(a), with the presence of PDEA. On one hand, this shows again that the existence of the macromolecule does not perturb the structure of the mixed solvent; on the other hand, it is a dilute solution and the conformation of PDEA will be strongly affected by the water-cosolvent interaction[34].

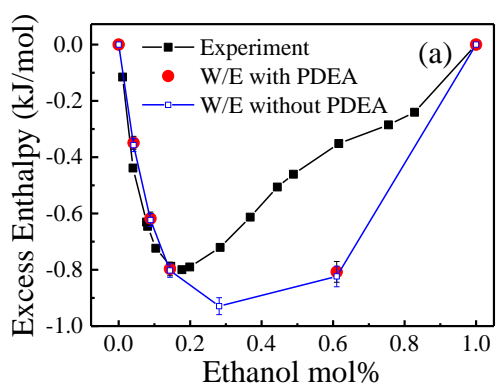

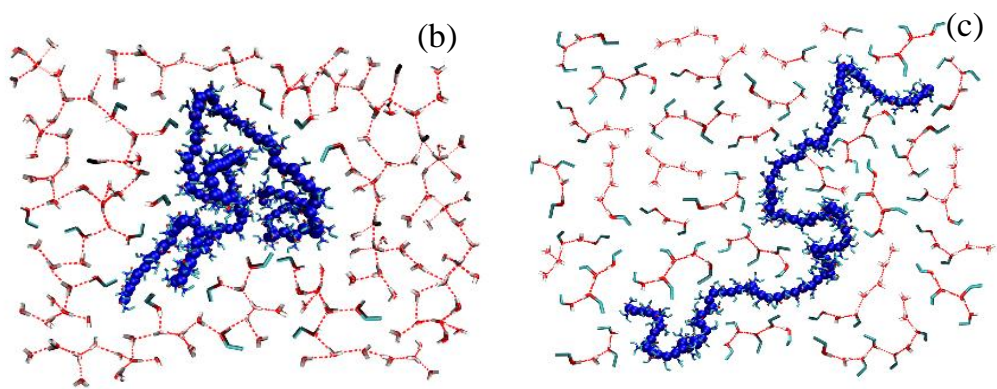

FIG. 4. Excess enthalpy of mixing water (W) and ethanol (E) and schematic illustration of the connonsolvency in PDEA ethanol-water system. (a) Excess enthalpy of water ethanol mixture from calorimetry experiment MD simulations in this work with and without the presence of PDEA. The experimental data was obtained from Ref.[31] with copy right granted. Excess enthalpy of mixing water-ethanol mixture was calculated by following the formulas of Zhong[35], and the interaction energy calculation between ethanol and water with the presence of PDEA were following the method of Dalgicdir[28]. (b) PDEA collapsed in water-rich region when it is higher than its LCST, where ethanol distributes in the tetrahedral structure of liquid water. (c) PDEA extended in water-poor region, where water complies with the zigzag structure of ethanol.

We can then check the entropy change. When ethanol is added into a PDEA aqueous solution, the total entropy change must be positive. However, because water-ethanol forms complexations and PDEA is a macromolecule, the real excess entropy in the cononsolvency process is smaller than that during random mixing[36].

Therefore, we propose the fourth view on cononsolvency, the strong water/cosolvent interaction, e. g., $\Delta H < 0$, is a necessary condition. PDEA is soluble in water via hydrogen bonding, while it dissolves in ethanol mainly via van der Waals interactions. When ethanol is mixed in the water-rich region, Some of ethanol molecules dissolve inside the tetrahedral structure of liquid water because of the strong water-ethanol attraction. Others compete to form hydrogen bonding to PDEA carbonyl side groups with water, the total number of hydrogen bondings with PDEA thus decreases, and the increase of van der Waals attraction between PDEA and ethanol in the hydration layer cannot compensate it. As a result, PDEA has to collapse to decrease its surface area in water-rich region at 27 ℃ (Fig. 4(b)). In the water-poor region, heterogeneities of water clusters distribute between the zigzag ethanol structure. PDEA re-swells because of the van der Waals attraction with ethyl groups of the ethanol molecules (Fig. 4(c)).

In conclusion, cononsolvency is a result of strong water-cosolvent attraction. Both preferential adsorption and complexation between solvent and cosolvent are just results of thermodynamic laws.

This work was funded by National Key Research and Development Program of China (2017 YFA0403703), National Nature Science Foundation of China (21474119 & 21674020), and UK-China Newton Project. We thank Dr. Tristan Youngs and Sam Callear in Nimrod beamline for setting the experiments, Professor van der Vegt and Dr. Angelina Folberth from Technische Universität Darmstadt for providing us the force field of PDEA, and the useful discussions with Prof. Zhong-Yuan Lu and Prof. Hu-Jun Qian from Jilin Univeristy, Prof. Hans Hasse and Dr. Edder Garcia Manzano from University of Kaiserslautern were also acknowledged.

T. Zuo and C. Ma contribute equally to the manuscript

\* Authors to whom correspondence should be addressed.

daniel.bowron@stfc.ac.uk

† Authors to whom correspondence should be addressed.

alan.soper@stfc.ac.uk

‡ Authors to whom correspondence should be addressed.

chenghe@ihep.ac.cn

# References

1. Brandrup J., Immergut E. H. & Grulke E. A. *Polymer Handbook Fourth Edition* (Wiley Press, 1999).
2. Miller-Chou B. A. & Koenig J. L. A review of polymer dissolution. *Prog. Polym. Sci.* **28**, 1223-1270 (2003).
3. Hore M. J. A., Hammouda B., Li Y. & Cheng H. Co-nonsolvency of poly(*N*-isopropylacrylamide) in deuterated water/ethanol mixtures. *Macromolecules* **46**, 7894-7901 (2013).
4. Amiya T., Hirokawa Y., Hirose Y., Li Y. & Tanaka T. Reentrant phase transition of *N*-isopropylacrylamide gels in mixed solvents. *J. Chem. Phys.* **86**, 2375-2379 (1987).
5. Tanaka F., Koga T. & Winnik F. M. Temperature-responsive polymers in mixed solvents: competitive hydrogen bonds cause cononsolvency. *Phys. Rev. Lett.* **101**, 028302 (2008).
6. Mukherji D., Marques C., M. & Kremer K. Collapse in two good solvents, swelling in two poor solvents: defying the laws of polymer solubility? *J. Phys. Condens. Matter* **30**, 024002 (2018).
7. Zhang G. & Wu C. Reentrant coil-to-globule-to-coil transition of a single linear homopolymer chain in a water/methanol mixture. *Phys. Rev. Lett.* **86**, 822-825 (2001).
8. Hao J., Cheng H., Butler P., Zhang L. & Han C. C. Origin of cononsolvency, based on the structure of tetrahydrofuran-water mixture. *J. Chem. Phys.* **132**, 154902 (2010).
9. Schild H. G., Muthukumar M. & Tirrell D. A. Cononsolvency in mixed aqueous solutions of poly(*N*-isopropylacrylamide). *Macromolecules* **24**, 948-952 (1991).
10. Tanaka F., Koga T. & Winnik F. M. Temperature-responsive polymers in mixed solvents: competitive hydrogen bonds cause cononsolvency. *Phys. Rev. Lett.* **101**, 028302 (2008).
11. Zhang G. & Wu C. The water/methanol complexation induced reentrant coil-to-globule-to-coil transition of individual homopolymer chains in extremely dilute solution. *J. Am. Chem. Soc.* **123**, 1376-1380 (2001).
12. Jia D., Zuo T., Rogers S., Cheng H., Hammouda B. & Han C. C. Re-entrance of poly(*N,N*-diethylacrylamide) in $D_2O$/d-ethanol mixture at 27 °C. *Macromolecules* **49**, 5152-5159 (2016).
13. Jia D., Muthukumar M., Cheng H., Han C. C. & Hammouda B. Concentration fluctuations near lower critical solution temperature in ternary aqueous solutions. *Macromolecules* **50**, 7291-7298 (2017).
14. Dudowicz J., Freed K. F. & Douglas J. F. Communication: Cosolvency and cononsolvency explained in terms of a Flory-Huggins type theory. *J. Chem. Phys.* **143**, 131101 (2015).
15. Dudowicz J., Freed K. F. & Douglas J. F. Solvation of polymers as mutual association. II. Basic thermodynamic properties. *J. Chem. Phys.* **138**, 164902 (2013).
16. Walter J., Sehrt J., Vrabec J. & Hasse H. Molecular dynamics and experimental study of conformation change of poly(*N*-isopropylacrylamide) hydrogels in mixtures of water and methanol. *J. Phys. Chem. B* **116**, 5251-5259 (2012).

17. Sørby M. H. Total neutron scattering. In: *Neutron scattering and other nuclear techniques for hydrogen in materials* (Fritzsche H, Huot J. & Fruchart D (Springer International Publishing, 2016).
18. Bowron D. T._, et al._ NIMROD: the near and intermediate range order diffractometer of the ISIS second target station. *Rev. Sci. Instrum.* **81**, 033905 (2010).
19. Head-Gordon T. & Johnson M. E. Tetrahedral structure or chains for liquid water. *Proc. Natl Acad. Sci. USA* **103**, 7973 (2006).
20. Soper A. K., Bruni F. & Ricci M. A. Site–site pair correlation functions of water from 25 to 400 °C: Revised analysis of new and old diffraction data. *J. Chem. Phys.* **106**, 247-254 (1997).
21. Bowron D. T., Finney J. L. & Soper A. K. Structural Investigation of Solute−Solute Interactions in Aqueous Solutions of Tertiary Butanol. *J. Phys. Chem. B* **102**, 3551-3563 (1998).
22. Soper A. K. & Finney J. L. Hydration of methanol in aqueous solution. *Phys. Rev. Lett.* **71**, 4346-4349 (1993).
23. Fidler J. & Rodger P. M. Solvation structure around aqueous alcohols. *J. Phys. Chem. B* **103**, 7695-7703 (1999).
24. Benmore C. J. & Loh Y. L. The structure of liquid ethanol: A neutron diffraction and molecular dynamics study. *J. Chem. Phys.* **112**, 5877-5883 (2000).
25. Yamaguchi T., Takamuku T. & Soper A. K. Neutron Diffraction Study on Microinhomogeneities in Ethanol-Water Mixtures. *J. Neutr. Res.* **13**, 129-133 (2005).
26. Fukai T., Shinyashiki N., Yagihara S., Kita R. & Tanaka F. Phase Behavior of Co-Nonsolvent Systems: Poly(N-isopropylacrylamide) in Mixed Solvents of Water and Methanol. *Langmuir* **34**, 3003-3009 (2018).
27. Mukherji D., Marques C. M. & Kremer K. Polymer collapse in miscible good solvents is a generic phenomenon driven by preferential adsorption. *Nat. Commun.* **5**, 4882 (2014).
28. Dalgicdir C., Rodríguez-Ropero F. & van der Vegt N. F. A. Computational calorimetry of PNIPAM cononsolvency in water/methanol mixtures. *J. Phys. Chem. B* **121**, 7741-7748 (2017).
29. Wang J._, et al._ Preferential adsorption of the additive is not a prerequisite for cononsolvency in water-rich mixtures. *Phys. Chem. Chem. Phys.* **19**, 30097-30106 (2017).
30. Dixit S., Crain J., Poon W. C. K., Finney J. L. & Soper A. K. Molecular segregation observed in a concentrated alcohol-water solution. *Nature* **416**, 829-832 (2002).
31. Kharin S. E., Byvaltsev Y. A. & Perelygin V. M. Über Mischungswärme von Ethanol und wasser. *Izv. Vyssh. Uchebn. Zaved. Pishch. Tekhnol.* **4**, 115-120 (1970).
32. Jorgensen W. L., Maxwell D. S. & Tirado-Rives J. Development and testing of the OPLS all-atom force field on conformational energetics and properties of organic liquids. *J. Am. Chem. Soc.* **118**, 11225-11236 (1996).
33. Rousseau R. W. & Fair J. R. *Handbook of separation process technology* (Wiley-IEEE Press, 1987).

34. Konstantinos K.*, et al.* Quantifying the Interactions in the Aggregation of Thermoresponsive Polymers: The Effect of Cononsolvency. *Macromolecular Rapid Communications* **37**, 420-425 (2016).
35. Zhong Y., Warren G. L. & Patel S. Thermodynamic and structural properties of methanol–water solutions using nonadditive interaction models. *J. Comput. Chem.* **29**, 1142-1152 (2008).
36. Soper A. K., Dougan L., Crain J. & Finney J. L. Excess entropy in alcohol-water solutions: a simple clustering explanation. *J. Phys. Chem. B* **110**, 3472-3476 (2006).

Supplemental Material for

# Water/Cosolvent Attraction Induced Phase Separation: a Molecular Picture of Cononsolvency

Taisen Zuo[1,2], Changli Ma[1,2], Guisheng Jiao[1,2], Zehua Han[1,2,3], Shiyan Xiao[4], Haojun Liang[4], Liang Hong[5], Daniel Bowron[6,*], Alan Soper[6,*], Charles C. Han[7], He Cheng[1,2,*]

[1] China Spallation Neutron Source (CSNS), Institute of High Energy Physics (IHEP), Chinese Academy of Science(CAS), Dongguan 523803, China

[2] Dongguan Institute of Neutron Science (DINS), Dongguan 523808, China

[3] University of Chinese Academy of Sciences, Beijing 100049, China

[4] CAS Key Laboratory of Soft Matter Chemistry, Collaborative Innovation Center of Chemistry for Energy Materials, Department of Polymer Science and Engineering, University of Science and Technology of China, Hefei, Anhui 230026, China.

[5] School of Physics and Astronomy &Institute of Natural Sciences, Shanghai Jiao Tong University, Shanghai 200240, China

[6] ISIS Facility, Rutherford Appleton Laboratory, Harwell Science and Innovation Campus, Didcot OX11 0QX, United Kingdom

[7] Institute for Advanced Study, Shenzhen University, Shenzhen, 508060, China

* Email: daniel.bowron@stfc.ac.uk

*Email: alan.soper@stfc.ac.uk

* Email: chenghe@ihep.ac.cn

Context:

Materials and Methods

Figs. S1 to S6 and Tables S1 and S2

References

**Materials and Methods**

**Materials**. PDEA was synthesized by reversible addition-fragmentation chain transfer polymerization (RAFT) introduced in our previous work[1]. The molar ratio of the reactants for DEA monomer: cumyl dithiobenzoate (CDB chain transfer agent): Azobisisobutyronitrile (AIBN initiator) was 600:1:0.2. In a typical reaction, 9.9500 g of DEA, 0.0355 g of CDB, and 0.0043 g of AIBN were dissolved in 10 mL DMF. The mixture was added in a polymerization tube. The tube was first frozen and thawed three times to remove oxygen, then put in an oil bath at 60 °C with a stirring speed at 6.7 Hz (6.7 revolutions per second or 400 rpm) for 5.6 hrs. After reaction, the monomer/polymer mixture was cooled to room temperature, dissolved in acetone (30 mL), and precipitated in a large amount of hexane. Finally, the product was dried in a vacuum oven at room temperature overnight. The relative weight-averaged molar mass of PDEA is 12000 g/mol (about 96 monomers) and the polydispersity index is 1.14 by Gel Permeation Chromatography (GPC). It was measured with polystyrene as standard and tetrahydrofuran as eluent.

**Neutron Total Scattering.** Sample preparing and neutron total scattering. Samples for total scattering were prepared by adding 0.0100 g of the as prepared PDEA to four water ethanol solutions with ethanol molar concentration of 4.1 mol%, 8.7 mol%, 14.2 mol% and 60.5 mol%, keeping PDEA monomer molar concentration 0.1 mol%. Corresponding deuterated and half deuterated samples were prepared by keeping the same molar ratio in PDEA $H_2O/D_2O$ and $C_2H_5OH/C_2D_5OD$ mixtures. Neutron total scattering experiments were carried out on the NIMROD diffractometer at the ISIS Pulsed Neutron Source (STFC Rutherford /Appleton Laboratory, Didcot, UK).[2,3] A simultaneous scattering vector (Q) range of 0.02 - 50 Å$^{-1}$ was achieved. The diffraction measurements were made on 1.4 cm$^3$ of the sample solutions. The samples were held in null scattering

Ti/Zr flat plate cells with a wall thickness of 1 mm, giving a sample thickness of 1 mm exposed to the beam that had a circular profile of 30 mm in diameter. The cells and the standard vanadium plate were loaded into the automatic sample changer. The temperature was maintained at 20 ℃, and measurements of each sample were made for approximately 4 hrs. Empty cell backgrounds and a 3 mm thick vanadium plate calibration standard were measured for an equivalent amount of time. Each raw scattering data was corrected for instrument and sample holder backgrounds, attenuation and multiple scattering using the instrument specific software Gudrun[4], the reduced scatterings were then normalized against to the known scattering of the vanadium calibration standard and converted to the interference differential scattering cross section $\frac{\partial \sigma(Q)}{\partial \Omega}$ vs Q for total scattering analysis as shown in Fig. S1, and total differential cross section ($\frac{\partial \Sigma(Q)}{\partial \Omega}$ vs Q for SANS analysis as shown in Fig. S2.

**Molecular dynamics Simulations.** The All atom molecular dynamics simulations were carried out on GROMACS 2016 package with TIP4PEW water and OPLS-AA PDEA and ethanol[5]. Simulation boxes were constructed with a single chain of atactic PDEA in each box with different composition of water/ethanol solutions. Numbers of water and ethanol were chosen to maintain the same atomic number density as in the neutron scattering experiments. The systems were simulated under constant pressure (1 bar) and constant temperature (293K) conditions using the Parrinello-Rahman borastat ($\tau_p$ = 2.0 ps) and V-rescale thermostat ($\tau_t$ = 0.1 ps). H-bonds were constrained by using the LINCS algorithm. Both van der Waals and coulomb radius cut-off were set as 1.5nm and the long range electrostatic interaction were calculated using the Particle Mesh Ewald (PME) method. An integration time step of 2 fs was used. All the MD simulations were conducted on the High-Performance Clusters (HPCs) of the National Supercomputer Shenzhen Centre and High-Performance Clusters at CSNS.

The initial conformations of PDEA were produced by the "build monopolymer" tool of the Material Studio Packages[6]. After relaxation of the initial conformation of PDEA in GROMACS, it was solvated and energy minimized using the steepest descent algorithm until convergence and then equilibrated in the canonical ensemble (NVT) for 100000 steps and then in the isothermal-isobaric ensemble (NPT) for 100000 steps before starting the production runs. More than 180 ns of data were accumulated for energy calculation in Fig. 4 and scattering profiles calculation in Fig. 1b and Fig. S1. Snapshots in this manuscript were rendered by the Visual Molecular Dynamics (VMD) [7].

**Fourier Transforms of MD simulations.** Neutron total scattering profiles of PDEA in fully deuterated water and ethanol as shown in Fig. 1b and Fig. S1 were compared with the Fourier Transforms of molecular dynamics simulations. The sizes of the simulation boxes are about 170 Å. According to the Periodic Boundary Conditions (PBCs), the smallest accessible scattering vector is about $2\pi/170 = 0.04$ Å$^{-1}$. Scattering profiles with scattering vector lower than 0.04 Å$^{-1}$ must be calculated without the PBCs. To keep us in a safe condition, all of the scattering curves with scattering vector from 0.06 Å$^{-1}$ to 50 Å$^{-1}$ (we call it diffraction curves) were calculated by the Debyer software package[8] with the PBCs, while the scattering curves with scattering vector from 0.02 Å$^{-1}$ to 1 Å$^{-1}$ (small angle scattering of PDEA in fully deuterated water) were fitted by the "gmx sans" tool (direct Fourier Transform software) of GROMACS without PBCs. The small angle scattering curves were shifted to overlap with the diffraction analysed results in the scattering vector ranging from 0.06 Å$^{-1}$ to 1 Å$^{-1}$. The resultant curves were then plotted with the experimental data as shown in Fig. 1b and Fig. S1. As for the half deuterated and fully protonated samples, only the diffraction curves calculated by the Debyer software package were fitted to the scattering curves. The Fourier Transforms of the samples with hydrogenous solvent are not so good at low Q due to the nonlinear inelastic scattering background from the hydrogen atoms which was not efficiently subtracted[9].

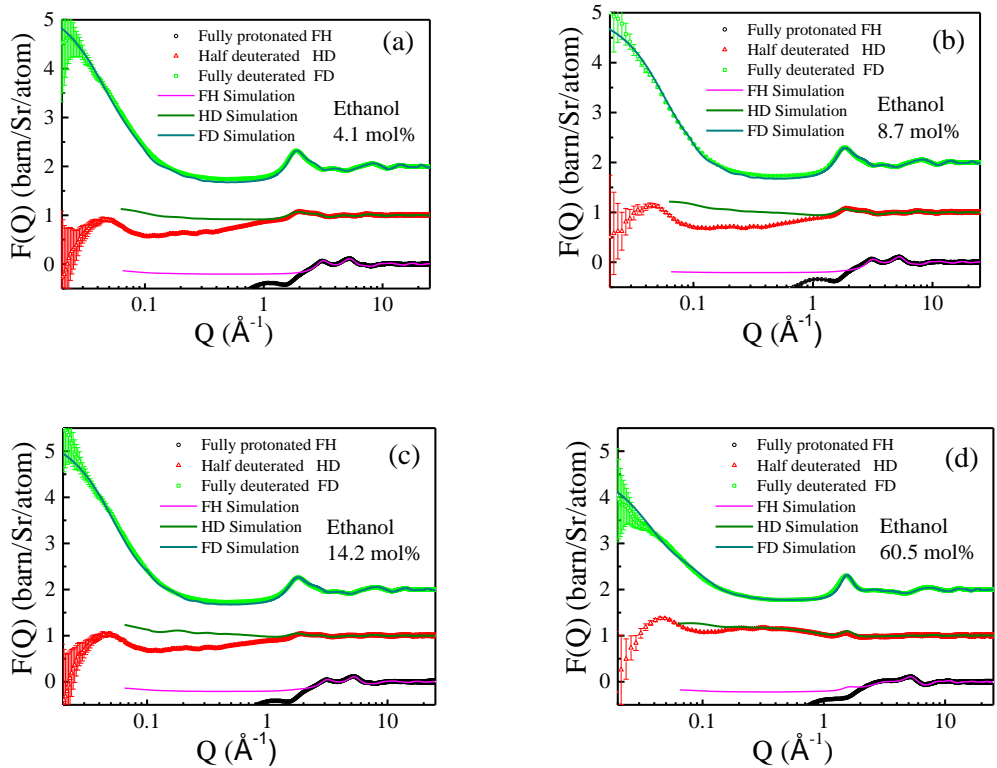

FIG. S1 (color online). Neutron total scattering profiles and the Fourier Transforms from molecular dynamic simulations. (a), (b), (c) and (d) are the neutron total scattering profiles of PDEA water ethanol solutions at 4.1 mol%, 8.7 mol%, 14.2 mol% and 60.5 mol% ethanol concentrations at 20 ℃, with water and ethanol molecules fully deuterated (FD), half deuterated (HD) and fully protonated (FH). The curves of FD and HD samples were shifted upward 2 and 1 barn/Sr/atom for clarity.

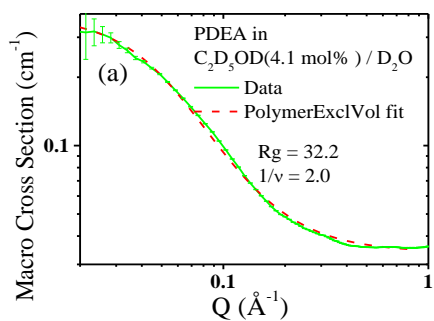
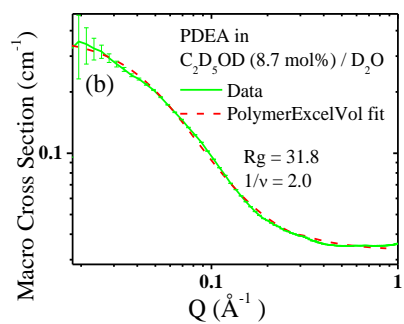

<, >

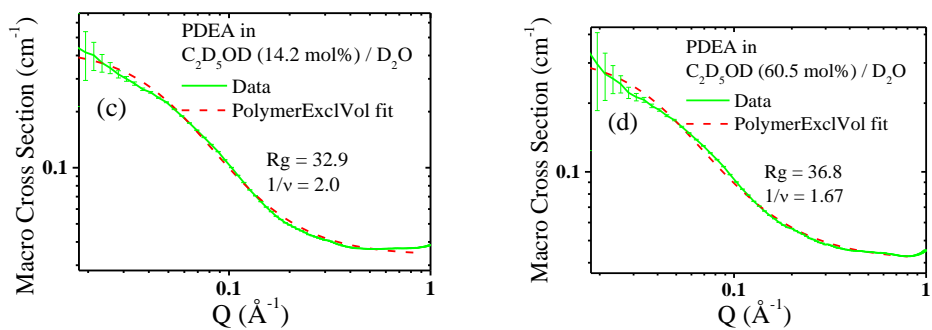

FIG. S2 (color online). Radius of gyration of PDEA fitted from the small angle neutron scattering (SANS) curves. The small angle scattering of PDEA in fully deuterated water-ethanol solution in Fig. S1 were exported by the data reduction software Gudrun with absolute intensity of macro cross section $\frac{\partial \Sigma}{\partial \Omega}$ (cm$^{-1}$). (a), (b), (c) and (d) are the best fits of the exported small angle scattering data of the PDEA in fully deuterated water ethanol solution at ethanol molar concentrations of 4.1 mol%, 8.7 mol%, 14.2 mol% and 60.5 mol%, respectively.

The fitting equation is the Polymer_Excl_Vol formula from the Data Analysis Package of NIST Center for Neutron Research,

$$\frac{\partial \Sigma}{\partial \Omega} = \phi \Delta \rho^2 N V_m P(Q) + bkg \qquad (1)$$

where $\phi$ is the macromolecule volume fraction, $\Delta \rho$ is the scattering contrast of the macromolecule and the solvent, $N$ is the degree of polymerization and $V_m$ is the volume of a monomer.

P(Q) in formula (1) is determined by equation

$$P(Q) = \frac{1}{\nu * U^{\frac{1}{2\nu}}} \gamma(\frac{1}{2\nu}, U) - \frac{1}{\nu * U^{\frac{1}{\nu}}} \gamma(\frac{1}{\nu}, U) \qquad (2)$$

Where

$$\gamma(x, U) = \int_0^U dt \exp(-t) t^{x-1}$$

$$U = \frac{Q^2 R_g^2 (2\nu + 1)(2\nu + 2)}{6}$$

Here, $\frac{1}{\nu}$ is the Porod exponent or factual dimension of the scattering object and $\nu$ is the scaling factor.

In ethanol-poor region, PDEA is a Gaussian coil without exclude volume, so $\nu$ = 2.0; while in ethanol-rich region, PDEA has excluded volume with $\nu$ = 1.67 according to the mean field theory. Therefore, the radius of gyration of PDEA can be fitted, as shown in Fig. S2.

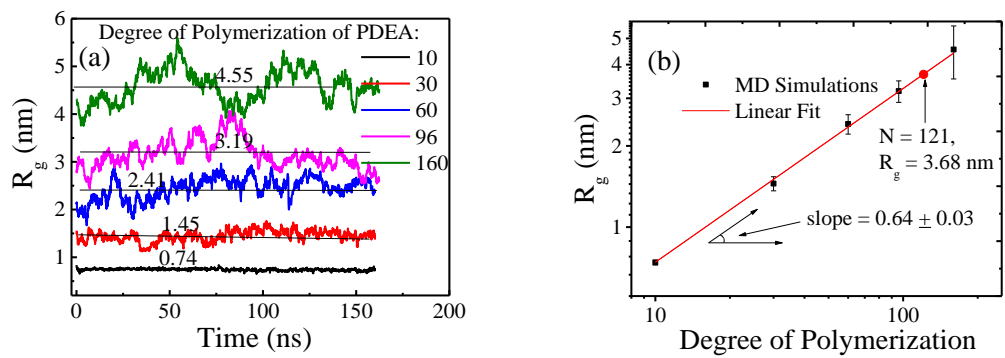

FIG. S3 (color online). Scaling of the radius of gyration $R_g$ of PDEA in the good solvent of ethanol. (a) Evolution of the $R_g$ of PDEA with 10, 30, 96 and 160 monomers in pure ethanol. (b) Scaling of $R_g$ Vs. the degree of polymerization.

According to the scaling law

$$R_g \propto N^{1/\nu} \qquad (3)$$

where, N is the degree of polymerization. The averaged $R_g$s of PDEA with degree of polymerization of 10, 30, 60, 96 and 160 from Fig. S3(a) were then fitted with equation (3). The fitted results showed that $\nu$ equals $0.64 \pm 0.03$. The real degree of polymerization of PDEA was determined to be 121, with $R_g$ equals 3.68 nm in a good solvent.

The resultant PDEA chains in the simulation box of Fig. 1(c) is

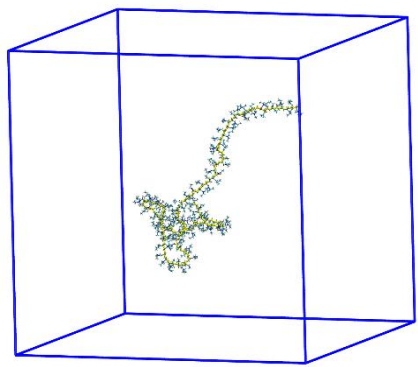

FIG. S4 (color online). The conformation of PDEA inside the simulation box of Fig. 1(c).

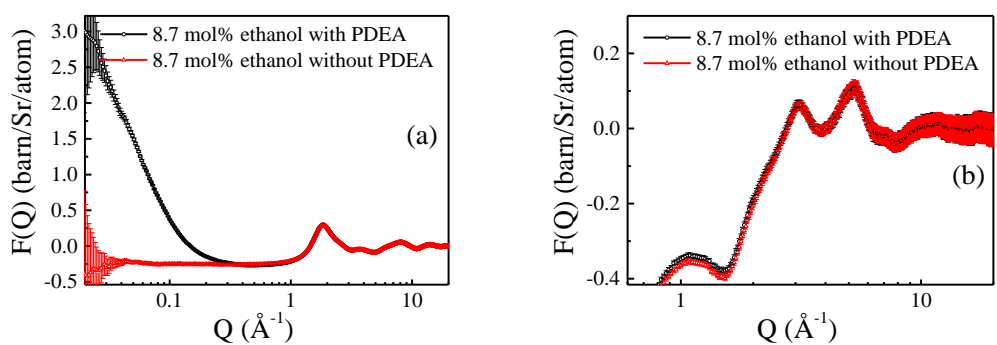

FIG. S5 (color online). Structure of water ethanol solution is not perturbed by the presence of PDEA (a) Scattering profiles of 8.7 mol% $C_2D_5OD$ in $D_2O$ with and without 0.1 mol% (monomer concentration) PDEA. (b) Scattering profiles of 8.7 mol% $C_2H_5OH$ in $H_2O$ with and without 0.1 mol% (monomer concentration) PDEA solvated.

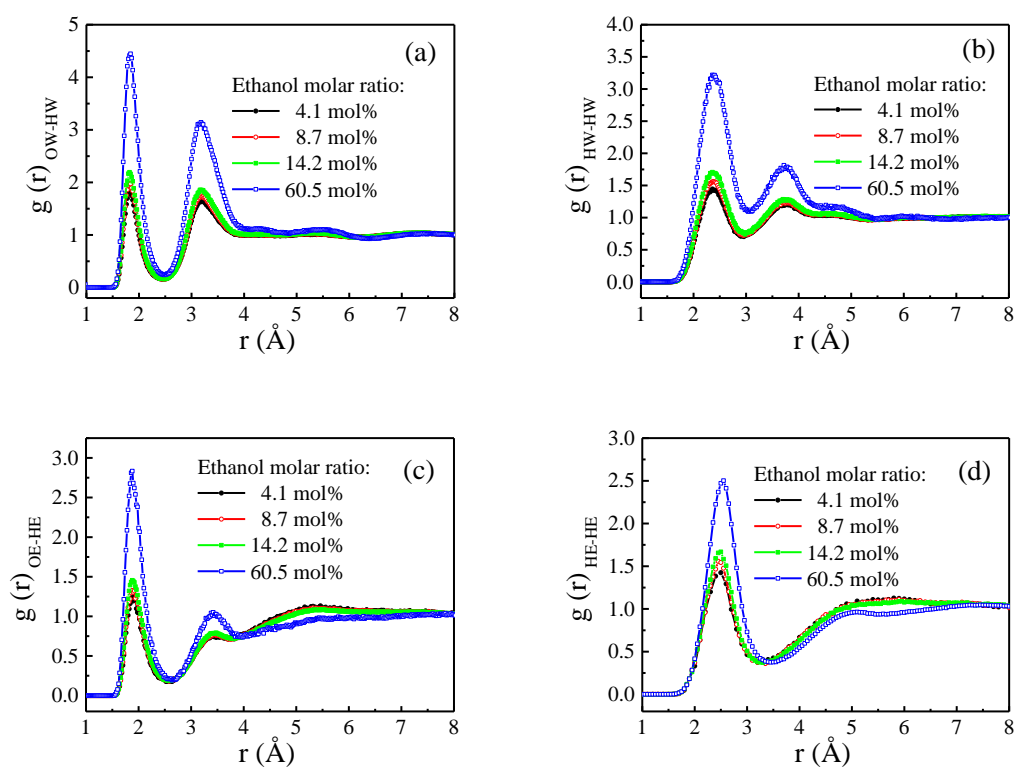

FIG. S6 (color online). Local structure of water (W) and ethanol (E). (a) $g(r)_{OW-HW}$ (OW means oxygen atoms of water, HW is the hydrogen atom of water); (b) $g(r)_{HW-HW}$; (c) $g(r)_{OE-HE}$, (OE means oxygen atom of ethanol, HE is the hydroxyl hydrogen atom of ethanol); (d) $g(r)_{HE-HE}$.

Table S1: Molar fraction and molecular numbers of PDEA water and ethanol in the MD simulations.

| Molar fraction of PDEA | Molar fraction of ethanol | Molar fraction of water | No. ethanol | No. water |
|---|---|---|---|---|
| 0.10% | 4.10% | 95.80% | 5212 | 121979 |
| 0.10% | 8.70% | 91.20% | 10478 | 109451 |
| 0.10% | 14.20% | 85.70% | 15644 | 94787 |
| 0.17% | 60.53% | 39.30% | 42917 | 27853 |

Table S2: Number of ethanol and water molecules in the MD simulations in calculating the excess enthalpy of mixing ethanol and water without PDEA in Fig. 4(a). Density of the water ethanol mixture were also shown.

| Molar fraction of ethanol | No. ethanol | No. water | Density |
|---|---|---|---|
| 4.1% | 5212 | 121979 | 0.983 |
| 8.7% | 10478 | 109451 | 0.970 |
| 14.2% | 15644 | 94787 | 0.955 |
| 27.4% | 33720 | 89400 | 0.920 |
| 60.6% | 42917 | 27853 | 0.857 |

**References:**

[1] D. Jia, T. Zuo, S. Rogers, H. Cheng, B. Hammouda, and C. C. Han, Macromolecules **49**, 5152 (2016).
[2] http://www.isis.stfc.ac.uk.
[3] D. T. Bowron *et al.*, Rev. Sci. Instrum. **81**, 033905 (2010).
[4] https://www.isis.stfc.ac.uk/Pages/Gudrun.aspx.
[5] J. T. Fern, D. J. Keffer, and W. V. Steele, J. Phys. Chem. B **111**, 13278 (2007).
[6] Materials Studio. Accelrys Software Inc., San Diego. Available from http://accelrys.com/products/materials-studio/.
[7] W. Humphrey, A. Dalke, and K. Schulten, J. Mol. Graph. **14**, 33 (1996).
[8] https://debyer.readthedocs.io/en/latest/.
[9] A. K. Soper, Molecular Physics **107**, 1667 (2009).